\UseRawInputEncoding
\documentclass[amsmath,amssymb,prx,superscriptaddress,reprint,showpacs,longbibliography]{revtex4-2}
\usepackage{times}
\usepackage{graphicx}
\usepackage{amsfonts}
\usepackage{amsmath, amsthm, amssymb}
\usepackage{dsfont}
\usepackage{color}
\usepackage{empheq}
\usepackage{standalone}
\usepackage[most]{tcolorbox}
\usepackage{physics}
\usepackage{bm}
\usepackage{amssymb,amsmath,amsfonts,amscd}
\usepackage{mathrsfs}
\usepackage{siunitx}
\usepackage{commath}
\usepackage{graphicx}      
\graphicspath{ {Images/} } 
\usepackage{braket}
\usepackage{bm}
\usepackage{empheq}
\usepackage[most]{tcolorbox}
\usepackage{physics}
\usepackage{placeins}

\usepackage{hyperref} 
\hypersetup{
    colorlinks=true,
    linkcolor=blue,    
    citecolor=blue,    
    urlcolor=blue,     
}

\begin{document}

\title{Evidence for strong electron correlations in a non-symmorphic Dirac semimetal}

\author{Yu-Te Hsu}
\email{yute.hsu@ru.nl}
\affiliation{High Field Magnet Laboratory (HFML-EMFL) and Institute for Molecules and Materials, Radboud University, Toernooiveld 7, 6525 ED Nijmegen, Netherlands}%

\author{Danil Prishchenko}
\affiliation{Department of Theoretical Physics and Applied Mathematics, Ural Federal University, 620002 Ekaterinburg, Russia}%

\author{Maarten Berben}
\author{Matija \v{C}ulo}
\author{Steffen Wiedmann}
\affiliation{High Field Magnet Laboratory (HFML-EMFL) and Institute for Molecules and Materials, Radboud University, Toernooiveld 7, 6525 ED Nijmegen, Netherlands}%

\author{Emily C. Hunter}
\affiliation{School of Physics and Astronomy, The University of Edinburgh, James Clerk Maxwell Building, Mayfield Road,
Edinburgh EH9 2TT, United Kingdom}

\author{Paul Tinnemans}
\affiliation{Department of Solid State Chemistry, Radboud University, Heyendaalseweg 135, 6525 AJ Nijmegen}%

\author{Tomohiro Takayama}
\affiliation{Max Planck Institute for Solid State Research, Heisenbergstrasse 1, 70569 Stuttgart, Germany}%

\author{Vladimir Mazurenko}
\affiliation{Department of Theoretical Physics and Applied Mathematics, Ural Federal University, 620002 Ekaterinburg, Russia}%

\author{Nigel E. Hussey}
\email{nigel.hussey@ru.nl}
\affiliation{High Field Magnet Laboratory (HFML-EMFL) and Institute for Molecules and Materials, Radboud University, Toernooiveld 7, 6525 ED Nijmegen, Netherlands}%
\affiliation{H. H. Wills Physics Laboratory, University of Bristol, Tyndall Avenue, Bristol BS8 1TL, United Kingdom}

\author{Robin S. Perry}
\email{robin.perry@ucl.ac.uk}
\affiliation{London Centre for Nanotechnology and Department of Physics and Astronomy, University College London, London WC1E 6BT, United Kingdom}%
\affiliation{ISIS Neutron and Muon Source, Rutherford Appleton Laboratory, Harwell, OX11 0QX, United Kingdom}%

\date{\today}

\begin{abstract}
Metallic iridium oxides (iridates) provide a fertile playground to explore new phenomena resulting from the interplay between topological protection, spin-orbit and electron-electron interactions. To date, however, few studies of the low energy electronic excitations exist due to the difficulty in synthesising crystals with sufficiently large carrier mean-free-paths. Here, we report the observation of Shubnikov-de Haas quantum oscillations in high-quality single crystals of monoclinic SrIrO$_3$ in magnetic fields up to 35~T. Analysis of the oscillations reveals a Fermi surface comprising multiple small pockets with effective masses up to 4.5 times larger than the calculated band mass. \textit{Ab-initio} calculations reveal robust linear band-crossings at the Brillouin zone boundary, due to its non-symmorphic symmetry, and overall we find good agreement between the angular dependence of the oscillations and the theoretical expectations. Further evidence of strong electron correlations is realized through the observation of signatures of non-Fermi liquid transport as well as a large Kadowaki-Woods ratio. These collective findings, coupled with knowledge of the evolution of the electronic state across the Ruddlesden-Popper iridate series, establishes monoclinic SrIrO$_3$ as a topological semimetal on the boundary of the Mott metal-insulator transition.
\end{abstract}

\maketitle

\section*{Introduction}
Band topology and strong electron correlations represent two of the most active research themes in quantum materials \cite{keimer2017}, with much attention now focused on the search for physics arising from their coexistence. To date, however, few examples of correlated topological materials are known to exist \cite{dzero2010, pezzini2018, kang2020, shi2021}. Among these, iridium oxides have emerged as one of the most promising material platforms on which to investigate the interplay between spin-orbit and electron-electron interactions \cite{tian2015, rau2016, cao2018}, due to the comparable energy scales of spin-orbit coupling ($\Lambda$), Coulomb repulsion ($U$) and electron bandwidth ($W$) \cite{moon2008}.

The Ruddlesden-Popper series Sr$_{n+1}$Ir$_n$O$_{3n+1}$, for example, exhibits a plethora of intriguing properties, such as spin-orbit-coupled Mottness \cite{kim2008}, pseudogap phenomenology \cite{kim2014, kim2015, delatorre2015, battisti2016}, odd-parity hidden order \cite{zhao2015}, and metal-insulator transitions \cite{ding2016, wu2013, biswas2014, groenendijk2017}. The spin-orbit-coupled $J_{\rm eff} = 1/2$ and $J_{\rm eff} = 3/2$ bands, resulting from crystal-field splitting of the $e_{\rm g}$ and $t_{\rm 2g}$ orbitals, are often treated as the starting point for their understanding \cite{rau2016}. In single-layer Sr$_2$IrO$_4$ ($n = 1$), the electron bandwidth of the half-filled $J_{\rm eff} = 1/2$ band is such that $U/W > 1$, resulting in a spin-orbit-coupled Mott insulator. With increasing $n$, electron hopping (and thereby $W$) increases while at the same time $U$ is reduced. Consequently, bilayer Sr$_3$Ir$_2$O$_7$ is found to be only weakly insulating \cite{ding2016} while the infinite-layer end member SrIrO$_3$ is itinerant, though its semimetallic ground state and low carrier concentration appear at odds with theoretical expectations of a metallic state with a half-filled $J_{\rm eff} = 1/2$ band and a large Fermi surface. Recently, the semimetallicity of SrIrO$_3$ was proposed to have a topological origin \cite{carter2012, zeb2012, takayama2019}.

SrIrO$_3$ crystallizes in two polymorphs: an ambient-pressure monoclinic phase and a high-pressure orthorhombic phase, which is also stabilized under epitaxial strain \cite{wu2013, biswas2014}. Three-dimensional (3D) Dirac points are created at the zone boundary of the non-symmorphic lattice by glide symmetry and are protected against gapping caused by spin-orbit coupling. This feature, coupled with their perceived proximity to a Mott transition, makes them attractive candidates for realizing a correlated topological semimetallic state. Until now, however, studies of their intrinsic electronic structure and properties have been restricted due to the fact that orthorhombic SrIrO$_3$ ($o$-SIO$_3$) can only be synthesized in thin-film form at ambient pressure \cite{moon2008, wu2013, biswas2014, nie2015, groenendijk2017, manca2018, sen2020}), whereas monoclinic SrIrO$_3$ ($m$-SIO$_3$), which can be synthesized in bulk form, has thus far remained largely unexplored. 

Here, we report the determination of the full Fermi surface of $m$-SIO$_3$ via the observation of Shubnikov-de Haas oscillations in high-quality single crystals. A number of small pockets are detected, in excellent agreement with first-principles density-functional theory (DFT) calculations. According to DFT, the electronic band structure of $m$-SIO$_3$ is found to be highly sensitive to atomic displacements, though the presence of linear (Dirac) band-crossings at the A- and M-points in the first Brillouin zone remains robust. Significantly, the experimentally determined effective masses are found to be substantially enhanced compared to the DFT band mass, signifying the presence of strong electron correlations.  We find a good overall agreement between the theoretically derived angular dependence of the quantum oscillation frequencies with the experimental observation, indicating that correlation effects play only a minor role in determining the Fermi surface topology of $m$-SIO$_3$. Further evidence for strong correlation effects is provided by the observation of a linear-in-temperature ($T$) component in the low-$T$ resistivity as well as an unusually large Kadowaki-Woods ratio. Overall, this study demonstrates that $m$-SIO$_3$ is a rare iridate topological semimetal displaying signatures of both a correlated Fermi liquid and a non-Fermi liquid ground state, possibly linked to its proximity to a Mott transition. The availability of high-quality single crystals and the sensitivity of its electronic structure to small perturbations make $m$-SIO$_3$ a promising platform on which to explore correlation-driven physics in a topological system.

\section*{Results and Discussions}
\subsection*{Shubnikov-de Haas oscillations}
\begin{figure}[hbtp!!!]
\includegraphics[width=1\linewidth]{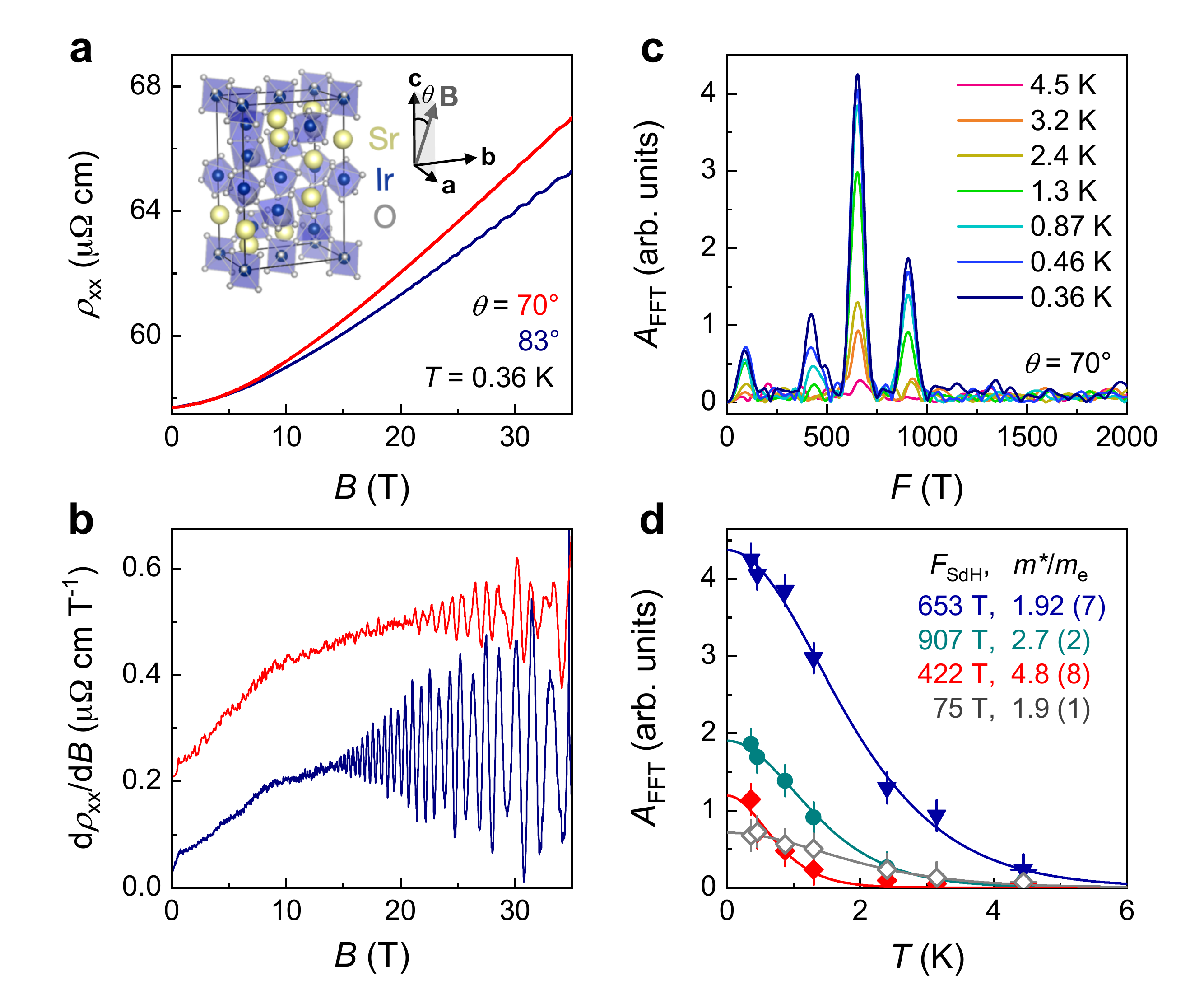}
\caption{\textbf{Shubnikov-de Haas oscillations in monoclinic SrIrO$_3$}.
\textbf{a} Electrical resistivity $\rho_{xx}$ measured in a magnetic field \textbf{B} up to 35~T at specified angles. Inset: crystal structure (in conventional unit cell) and experimental alignment. \textbf{B} is rotated within the $a$-$c$ plane, with $\theta$ denoting the angle between \textbf{B} and the $c$-axis [001]. 
\textbf{b} Field-derivative of the data shown in \textbf{a} using the same colour coding. $\theta = 70^{\circ}$ data are shifted vertically for clarity. 
\textbf{c} Fourier spectrum of the SdH oscillations measured at $\theta = 70^{\circ}$ and indicated temperatures, after a polynomial background is subtracted. Four distinct peaks at $F_{\rm SdH}$ = 75, 422, 653, and 907~T are resolved using a field window of 18~T $< B <$ 35~T.
\textbf{d} Extraction of the effective masses $m^*$ through fitting the oscillation amplitude $A_{\rm FFT}(T)$ using the Lifshitz-Kosevich expression (see text). Error bars in $A_{\rm FFT}$ reflect the noise floor of the Fourier spectra shown in \textbf{c}.}
\label{SdH}
\end{figure}

Figure 1\textbf{a} shows field-dependent resistivity $\rho_{xx}(B)$ curves measured at $T$ = 0.36~K and two angles $\theta = 70^{\circ}$ and 83$^{\circ}$ relative to the [001]-axis (see inset to Fig.~\ref{SdH}\textbf{a} for experimental alignment). Prominent Shubnikov-de Haas (SdH) oscillations emerge above 20~T (most evident in the $\theta = 83^{\circ}$ trace). By plotting the field-derivative d$\rho_{xx}$/d$B$ in Fig. 1\textbf{b}, the full oscillatory signals become visible for both angles. The oscillation waveform at $\theta = 83^{\circ}$, which consists of four oscillatory components as revealed by the fast Fourier transform (FFT) spectrum, is dominated by the $F_{\rm SdH}$ = 650 T component. At $\theta = 70^{\circ}$, all four constituent frequencies, with distinct peaks at $F_{\rm SdH}$ = 75, 422, 653, and 907 T, can be tracked over a range of temperatures, as shown in Fig.~1\textbf{c}, thereby enabling a more reliable determination of the corresponding quasiparticle effective masses $m^*$.
Each frequency $F_{\rm SdH}$ is related to an extremal cross-sectional area ($A_k$) of the Fermi surface perpendicular to the field direction via the Onsager relation $F_{\rm SdH} = (\hbar/2\rm{\pi e})$$A_k$. 
By fitting the $T$-dependence of the oscillation amplitude $A_{\rm FFT}(T)$ at $\theta =  70^{\circ}$ using the Lifshitz-Kosevich expression for the thermal damping term \cite{shoenberg}
    $R_T = X/\sinh X,$
where $X$ = 14.69 $m^*T/B$, we obtain $m^*$ values associated with each frequency ranging from 1.9 to 4.8~$m_{\rm e}$. Such high masses are exceptional for a Dirac semimetal. 

\subsection*{DFT calculations}

\begin{figure*}[hbtp!!!]
\includegraphics[width=1\linewidth]{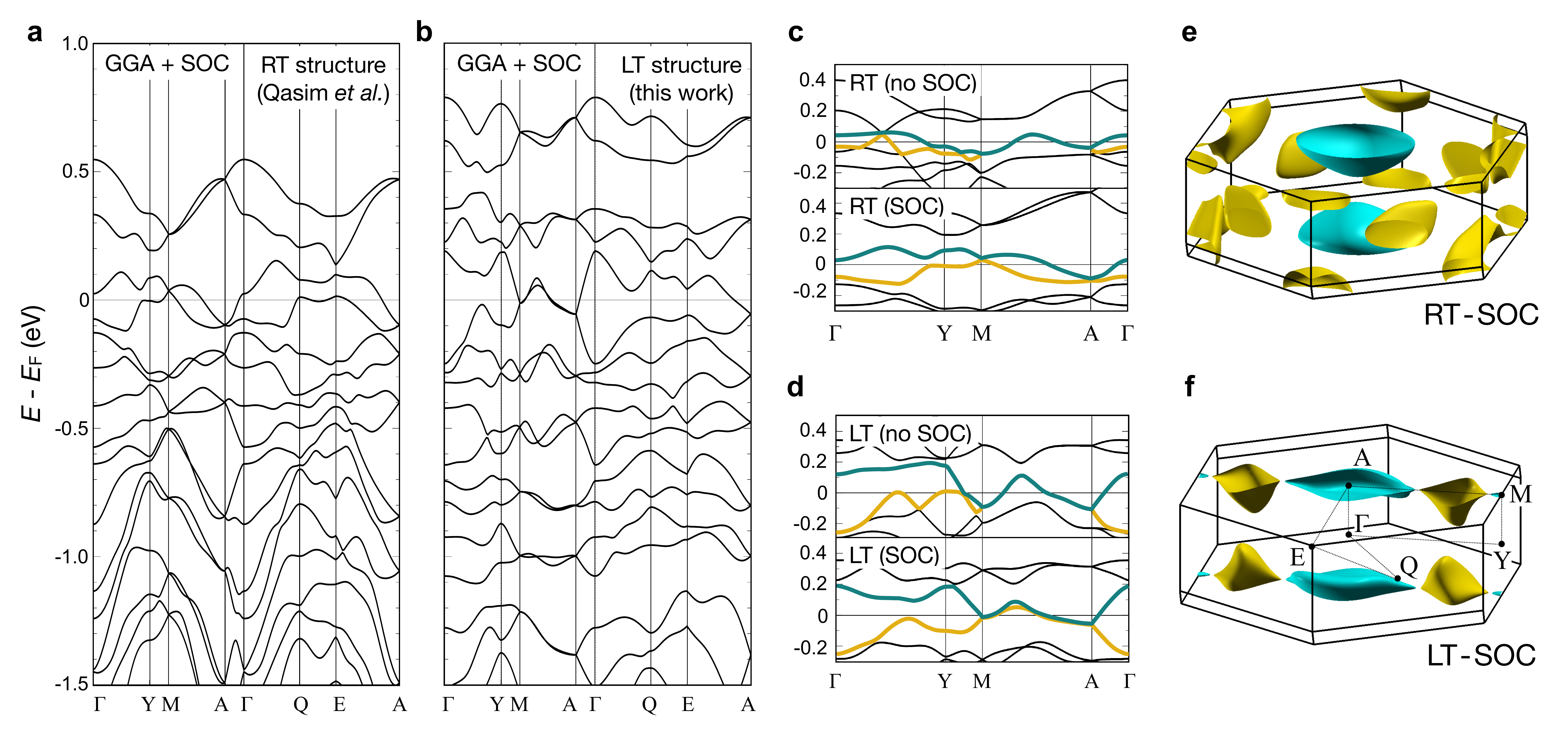}
\caption{\textbf{Electronic structure of $m$-SIO$_3$ calculated by density-functional theory}.
\textbf{a, b} Band structures obtained using room-temperature (RT) lattice parameters reported in \cite{qasim2013} and low-temperature (LT) lattice parameters found at 13~K on our crystals. The high-symmetry points in the first Brillouin zone are labelled in \textbf{f}. 
\textbf{c, d} Comparison of the RT and LT band structures with and without SOC near the Fermi energy $\epsilon_{\rm F}$. Bands that give rise to electron and hole Fermi pockets are highlighted in cyan and yellow, respectively. Note that linear band-crossings at the M- and A-points are found in both GGA+SOC results despite the overall differences.
\textbf{e, f} Corresponding Fermi surface models for the RT and LT band structures (GGA+SOC). Fermi pockets are colour-coded in accordance with \textbf{c, d}.}
\label{DFT}
\end{figure*}

In order to understand the origin of the observed SdH oscillations, we performed DFT band structure calculations using the generalized gradient approximation with effects of spin-orbit coupling included (GGA+SOC). Figure~\ref{DFT} shows the calculated band structures using the structural parameters reported in \cite{qasim2013}, refined at room temperature (RT), and those found on our own single crystals, refined at low temperature (LT; 13~K) (see Supplementary Methods). The RT band structure agrees well with a previous report \cite{takayama2019}, with linear band-crossings near the Fermi energy $\epsilon_{\rm F}$ at the M- and A-points. The LT band structure, however, reveals a markedly different picture, though notably the linear band-crossings at M- and A-points are still preserved. 

The effects of SOC on each band structure are illustrated in Fig.~\ref{DFT}\textbf{c,d}. The inclusion of SOC leads to a separation of the low-lying bands near $\epsilon_{\rm F}$ by up to $\sim$ 0.3~eV, while the degeneracy at the M- and A-points remains unaffected. 
The resultant Fermi surfaces from the two GGA+SOC band structures, shown in Fig.~\ref{DFT}\textbf{e,f}, show a number of notable differences which we attribute to the extreme sensitivity of the electronic structure of $m$-SIO$_3$ with respect to the precise atomic positions -- known to be a challenging issue for iridates \cite{sen2020, thiel2020} -- rather than to a structural phase transition. 
The different types of samples used for the structural refinements -- polycrystalline powder in \cite{qasim2013} and crushed single crystals in our study -- can cause minute structural differences and, in turn, lead to differences in the corresponding band structures.
Nevertheless, the linear band-crossings near $\epsilon_{\rm F}$ at the M- and A-points are found to be a robust feature independent of such structural details, confirming their protection by the underlying non-symmorphic lattice symmetry.

\begin{figure}[hbtp!!!]
\includegraphics[width=1\linewidth]{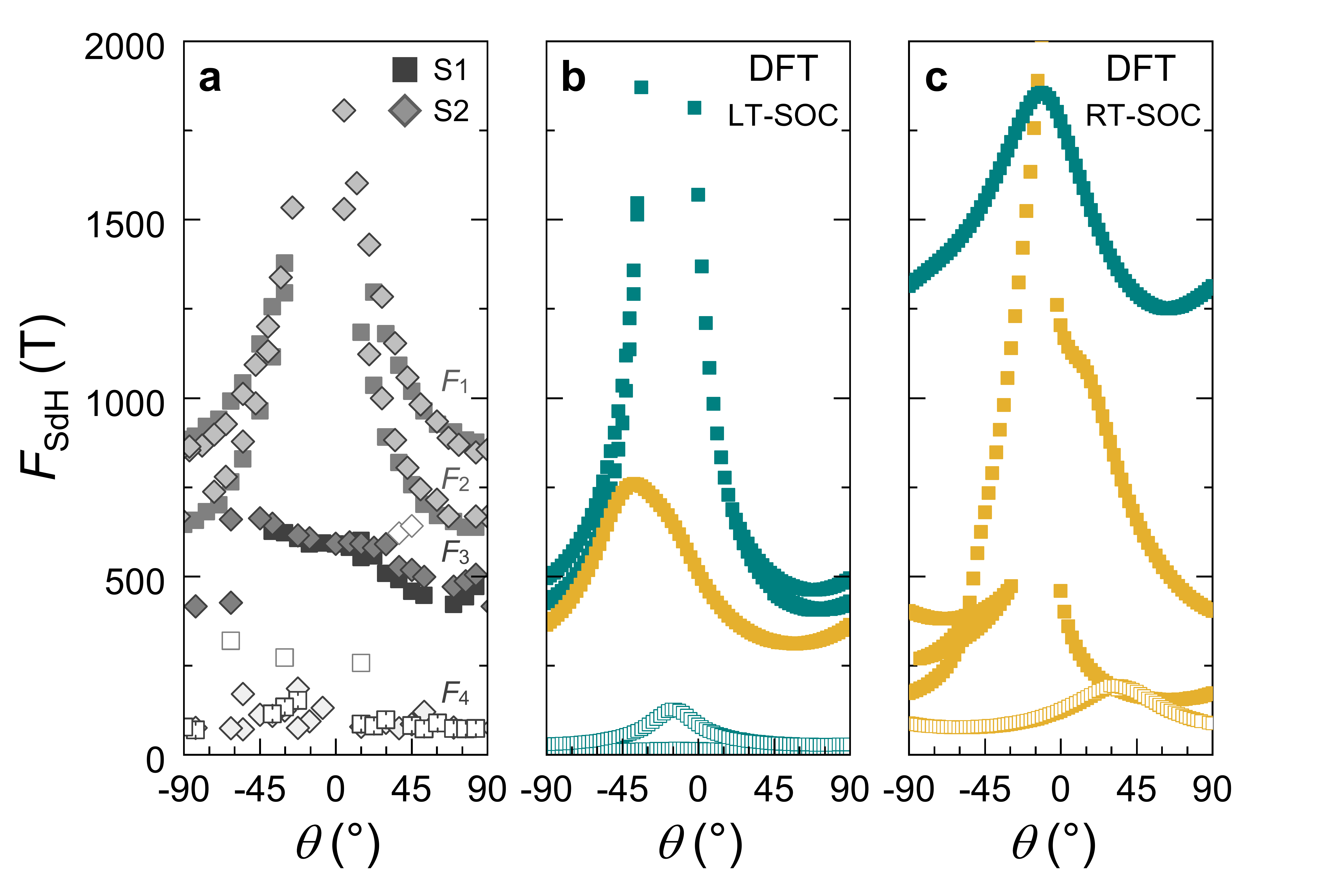}
\caption{\textbf{Angular dependence of the observed and calculated quantum oscillation frequencies}.
\textbf{a} $F_{\rm SdH}(\theta)$ measured in two samples, S1 and S2, cut from different crystals of the same growth batch. Four distinct branches can be identified, labelled as $F_{1}$ to $F_4$. 
\textbf{b, c} $F_{\rm SdH}(\theta)$ expected from the LT and RT Fermi surface models as shown in Fig.~\ref{DFT}, with the electron and hole-pockets colour-coded accordingly. Note that the $y$-axis for all panels is the same.}
\label{angular}
\end{figure}

In order to identify the appropriate Fermi surface model from the DFT calculations, we compare in Fig.~\ref{angular} the angular dependence of the observed quantum oscillation frequencies $F_{\rm SdH}(\theta)$ with the predictions emerging from each model. As shown in Fig.~\ref{angular}\textbf{a}, the four distinct frequency branches disperse differently with angle $\theta$ and range from 50~T to 2000~T.
Overall, we find a better agreement between the experiment with the LT-SOC model (referred to as the LT model hereafter), which reproduces the close tracking between $F_1$ and $F_2$, the rapid increase in $F_{\rm SdH}$ as $\theta \rightarrow 0$, as well as the moderate variation of $F_3$ with respect to $\theta$. This observation implies all the observed pockets originate from the metallic bulk. While the apparent divergence of frequencies $F_{1, 2}$ as \textbf{B} approaches [001] suggests a possible surface state originating in the (100) plane, we can exclude such a possibility by examining its precise angle dependence.
For a surface orbit, the SdH frequency is expected to follow a $1/\cos\phi$ dependence as \textbf{B} is tilted at an angle $\phi$ away from the surface normal in which the orbit is hosted. 
As shown in Supplementary Fig. 3, by plotting $F_{1,2}\cos\phi$ as a function of $\phi$, where $\phi = 90^{\circ} - \theta$ denotes the tilt angle between \textbf{B} and [100], we find $F_{1,2}\cos\phi$ deviates strongly from a constant value as $\phi$ exceeds 30$^{\circ}$, inconsistent with expectations for a surface origin.
We therefore conclude that there is no surface contribution to the observed SdH frequencies. 
We note that the LT model underestimates the $F_{\rm SdH}$ for the $F_{1,2}$ branch as $\theta \rightarrow$ 90$^{\circ}$, indicating the electron pocket at M-point is less elongated in reality than that illustrated in Fig.~\ref{DFT}\textbf{f}, possibly due to subtle atomic displacements and/or correlation effects.

Nevertheless, despite the technical challenges associated with DFT calculations for this complex oxide, the overall agreement between the experimentally observed SdH frequencies and the theoretical expectations from the LT model suggests that the full Fermi surface of $m$-SIO$_3$ has been determined by our study. The Dirac points lie $\approx$ 15 and 50~meV below the Fermi level at the M- and A-points, respectively, at which the band dispersion remains linear (see Fig.~\ref{DFT}\textbf{d}).
In order to identify the correct FS model from the DFT calculations, we compare in Fig.~\ref{angular} the angular dependence of the observed SdH frequencies $F_{\rm SdH}(\theta)$ with the predictions emerging from each model. As shown in Fig.~\ref{angular}(a), the four distinct frequency branches disperse differently with angle $\theta$ and range from 50~T to 2000~T. While both models predict four branches at most angles and an electron-pocket associated with the highest frequency branch (Fig.~\ref{angular}(b, c)), we find better overall agreement with the LT-SOC model (referred to as the LT model hereafter). Importantly, the close tracking between $F_1$ and $F_2$, the rapid increase in $F_{\rm SdH}$ as $\theta \rightarrow 0$, and the moderate variation of $F_3$ with respect to $\theta$ are all well-reproduced in the LT model.

The rapid increase of $F_{1,2}$ as \textbf{B} is tilted away from [100] may suggest a surface state located within the (100) plane. For such an orbit, the SdH frequency should follow a $1/\cos\phi$ dependence as \textbf{B} is tilted away from the surface. We examine this possibility by plotting $F_{1,2}\cos\phi$ versus $\phi$, where $\phi = 90^{\circ} - \theta$ denotes the tilt angle between \textbf{B} and [100], as shown in Supplemental Material Fig.~S2. We find $F_{1,2}\cos\phi$ deviates strongly from a constant value as $\phi$ exceeds 30$^{\circ}$, inconsistent with expectations for a surface state. 
We therefore conclude that all the observed pockets arise from the metallic bulk. Despite the technical challenges associated with DFT calculations for this complex oxide, the remarkable agreement in the experimentally observed angular dependence of SdH frequencies and the theoretical expectations from the LT model suggests that the full FS of $m$-SIO$_3$ has been accurately determined by our study.

\subsection*{Signatures of strong electronic correlations}
\begin{table*}[hbtp!!!]
\begin{ruledtabular}
\caption{\textbf{Fermi surface parameters found by experiment and DFT calculations (LT model).} $l$, $m^*$, and $m_{\rm b}$ represent quasiparticle mean free path, effective mass, and DFT band mass, respectively. All parameters are extracted at $\theta = 70^{\circ}$. 
Note that a reliable extraction of $l$ associated with $F_4$ is precluded by the dominating influence of the higher-frequency oscillations, and the parameters marked by $\dagger$ are an average over the multiple small orbits ($<$ 30~T) predicted by the DFT model that cannot be resolved experimentally.}
\begin{tabular}{ccccccc}
 & $F_{\rm SdH}$ (T) & $F_{\rm DFT}$ (T) & $l$ (nm) & $m^*$ ($m_{\rm e}$) & $m_{\rm b} (m_{\rm e})$ & $ m^*/m_{\rm b}$\\
\colrule
$F_1$ & 907 & 462 & 30 $\pm$ 2 & 2.81 $\pm$ 0.03 & 1.46 & 1.9\\
$F_2$ & 653 & 410 & 39 $\pm$ 3 & 1.90 $\pm$ 0.03& 1.36 & 1.4\\
$F_3$ & 422 & 318 & 26 $\pm$ 3 & 5.8 $\pm$ 0.7 & 1.07 & 5.4\\
$F_4$ & 75 & 15$^{\dagger}$ & - & 2.0 $\pm$ 0.4 & 1.04$^{\dagger}$ & 1.9$^{\dagger}$
\end{tabular}
\label{FSparameters}
\end{ruledtabular}
\end{table*}

\begin{figure*}[bthp!!!]
\centering
\includegraphics[width=1\linewidth]{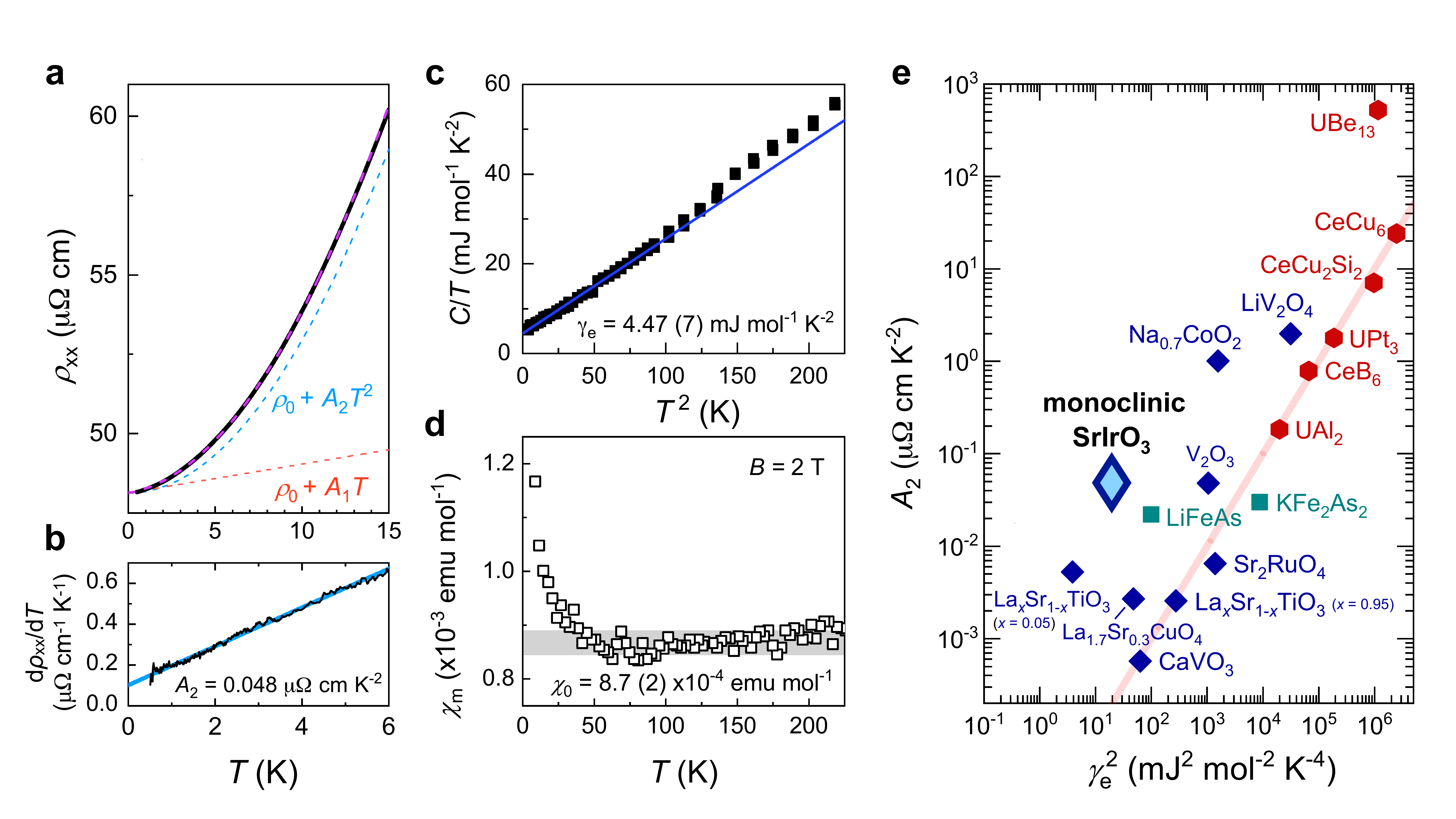}
\caption{\textbf{Non-Fermi liquid transport and Kadowaki-Woods ratio of $m$-SIO$_3$.}
\textbf{a} Zero-field resistivity $\rho_{xx}(T)$ below 15~K and, \textbf{b} its derivative d$\rho_{xx}$/d$T$ below 6~K, revealing a finite $T$-linear component in $\rho_{xx}(T)$ at low $T$. Purple dashed line in \textbf{a} is a fit to the experimental data using the functional form: $\rho_{xx}(T) = \rho_0 + A_1 T + A_2 T^2$, with the $T$-linear and $T^2$-components shown in red and blue, respectively. 
\textbf{c} Specific heat plotted as $C/T$ versus $T^2$ and fitted with $C/T = \gamma_{\rm 0} + \beta T^2$ below 10~K (blue line).
\textbf{d} Magnetic susceptibility $\chi_{\rm m}$ measured with $B$ = 2~T applied along [001]. $\chi_{\rm m}$ is largely $T$-independent down to $\approx$ 30K, with a constant $\chi_0$ as marked by the grey band. Below 30~K, $\chi_{\rm m}$ shows a small upturn, which we attribute to an impurity contribution that follows a Curie-Weiss behaviour. 
\textbf{e} Transport coefficient $A_2$ plotted against $\gamma_0^2$, known as the Kadowaki-Woods plot, for selected correlated oxides (diamonds), iron pnictides (hexagons), heavy fermion materials (circles) (\cite{hussey2005, jacko2009, cavanagh2015} and references therein), and monoclinic SrIrO$_3$ (elongated diamond). Red line corresponds to $A_2/\gamma_0^2$ = 10~$\mu\Omega$~cm~mol$^2$~K$^2$~mJ$^{-6}$, known to describe many heavy fermion materials well. 
}
\label{NFL}
\end{figure*}

Having established the appropriate Fermi surface model, we proceed to quantify the extent of electron correlation in $m$-SIO$_3$ by comparing the measured effective masses $m^*$ with the calculated band masses $m_{\rm b}$. 
Here we use the definition $m_{\rm b} = \hbar^2(dA_k/dE)|_{\epsilon_{\rm F}}$, namely the change of area enclosed by the cyclotron orbit in $k$-space with respect to the change in energy at the Fermi level, to describe the band mass of quasiparticles with an arbitrary dispersion $\epsilon(k)$.
As shown in Table~1, we find a substantial mass renormalization factor $m^*/m_{\rm b}$ for all frequencies, ranging from 1.4 to 4.5, signifying strong electron correlations. We note that $m^*/m_{\rm b}$ is larger for the hole pocket ($F_3$) compared to the electron pockets with linear band-crossings, possibly reflecting the different impact of correlation effects on the conventional and Dirac-like bands.

The inclusion of $U$ up to 3~eV in our calculations leads to a considerable modification in the Fermi surface topology, as shown in Supplementary Fig. 4. The best agreement with experiment, however, is found when $U$ = 0, indicating that the influence of electron correlations cannot be effectively captured for complex iridates using the GGA+SOC+$U$ approach \cite{nie2015}. A more accurate accounting of correlation effects will likely require the implementation of computationally demanding dynamical mean-field theory calculations, which will be the subject of a future investigation. Nevertheless, the remarkable agreement between the LT model and experiment demonstrates that correlation effects play only a minor role in determining the Fermi surface topology, as found for related systems such as $o$-SIO$_3$ \cite{nie2015} and the topological semimetal ZrSiS \cite{mueller2020}.

The anomalous nature of the electronic state in $m$-SIO$_3$ is revealed through its transport and thermodynamic properties at low temperatures. $\rho_{xx}(T)$ below 15~K follows an unusual $T + T^2$-dependence, as revealed by the finite intercept in d$\rho_{xx}$/d$T$ at $T = 0$ (Fig.~\ref{NFL}\textbf{a, b}). The $T$-linear term $A_1$ = 0.1 $\mu\Omega$~cm~K$^{-1}$, though small, persists down to the lowest temperatures, possibly indicating the presence of low-energy critical fluctuations. By contrast, the specific heat and magnetic susceptibility of $m$-SIO$_3$, shown in Fig.~\ref{NFL}\textbf{c, d}, are more characteristic of a paramagnetic Fermi-liquid state. 
Using the coefficient of the $T^2$ term $A_2$ = 0.048~$\mu\Omega$~cm~K$^{-2}$ and the electronic specific heat coefficient $\gamma_0$ = 4.47~(7)~mJ~mol$^{-1}$~K$^{-2}$, the corresponding Kadowaki-Woods ratio ($R_{\rm KW}$) is found to be $A_2/\gamma_0^2 \approx$ 2400~$\mu\Omega$~cm~K$^2$~mol$^2$~J$^{-2}$ or\\ $A_2/\gamma_{V}^2 \approx$ 3.5~$\mu\Omega$~cm~K$^2$~cm$^6$~J$^{-2}$, where $\gamma_V$ is the volume form of $\gamma_0$ \cite{hussey2005}. 
These values, though high relative to other correlated metals \cite{hussey2005, jacko2009, cavanagh2015}, are nonetheless consistent with theoretical estimates derived for a 3D Fermi surface \cite{hussey2005} (see Supplementary Discussion for details), implying that electron-electron scattering dominates the transport behaviour at low $T$.
Recently, it has been shown that the strength of electron interactions correlates with $R_{\rm KW}$ in the most general case \cite{cavanagh2015}, and overall consistency in our analysis indicates that electron-electron scattering is responsible for the large $R_{\rm KW}$ and the highly renormalized $m^*$ in $m$-SIO$_3$, possibly due to the SOC-renormalized bandwidth and/or proximity to a quantum critical point \cite{cao2007, groenendijk2020}.

The topological character of the electronic bands in $m$-SIO$_3$ is examined by extracting the phase factor of the SdH oscillations (see Supplementary Fig.~5). While we find a non-trivial phase factor $\approx\rm{\pi}$, as expected for a topologically non-trivial band, recent theoretical findings have suggested that the phase factor in quantum oscillations can comprise of multiple contributions and therefore cannot be considered as a conclusive evidence for a non-trivial band topology \cite{alexandradinata2018}. 
Nonetheless, we emphasise that the topological non-trivial character for the electron band in $m$-SIO$_3$ is further supported by the good agreement between the DFT calculations -- with their topologically-protected Dirac points at M and A -- and the observed quantum oscillation angular dependence.

Recently, transport signatures of Dirac quasiparticles have been reported in a closely-related iridate CaIrO$_3$ \cite{fujioka2019, yamada2019}. The resultant Fermi surface areas and effective masses, however, were found to be small ($F_{\rm SdH}$ = 3.2 $-$ 11.2~T and $m^*$ = 0.12 $-$ 0.31 $m_{\rm e}$), suggesting that the Fermi level in CaIrO$_3$ is much closer to the Dirac point and that electron correlations are considerably weaker than in $m$-SIO$_3$. A recent photoemission experiment on $o$-SIO$_3$ thin films, on the other hand, found an electron-like band with linear dispersion and parabolic hole-like bands with heavy quasiparticle masses $m^* = 2.4-6.0$ $m_{\rm e}$ \cite{nie2015}, largely in agreement with our findings on $m$-SIO$_3$. 
Collectively, these findings identify SIO$_3$ as a rare example of a topological semimetal with enhanced electron correlations, possibly induced through proximity to a Mott instability \cite{moon2008}.
The extreme sensitivity of its low-energy electronic structure to atomic displacement, highlighted by the DFT calculations, further identifies $m$-SIO$_3$ as a viable platform on which to explore the Mott transition in a topological semimetal, by tuning the relative strength of the electronic bandwidth and Coulomb repulsion via doping or via structural tuning parameters such as hydrostatic pressure or uniaxial strain.

\section*{Methods}
\subsection*{Crystal growth and characterization}
Single crystals of $m$-SIO$_3$ were grown using the flux method described in Ref.~\cite{hunter}. 
Low-temperature structural refinement using X-ray diffraction (XRD) was performed at 13~K (see Supplementary Methods for details). 

\subsection*{Measurements}
Electrical resistivity measurements were performed on Hall bars of ($600 \times 100 \times 60$) $\mu$m$^3$ in size, cut from pristine crystals, with an ac current of 3~mA applied along the bar identified as the crystallographic axis [100] by Laue XRD. 
Magnetic fields up to 35~T were generated using a Bitter magnet at the High Field Magnet Laboratory (HFML) in Nijmegen, the Netherlands. 
Zero-field resistivity measurements were performed using a Cryogenic Free Measurement System by Cryogenic Limited; specific heat and magnetic susceptibility measurements were performed using a Physical Properties Measurement System and a Magnetic Properties Measurement System XL, respectively, by Quantum Design Inc.

\subsection*{Band structure calculations}
DFT band structure calculations were carried out using {\it ab-initio} VASP code~\cite{vasp1} with energy cut-off set to 400~eV and Brillouin zone sampled by the $9 \times 9 \times 5$ Monkhorst-Pack $k$-point mesh. We used WANNIER90 package \cite{wannier90} to construct the corresponding tight-binding (TB) Hamiltonian and ensured that the resulting VASP and TB electronic band structures are identical near the Fermi level. The visualization of Fermi surfaces was done using XCrysDen code \cite{xcrysden} and the oscillation frequency expected from the DFT Fermi surface model was calculated using the SKEAF code \cite{rourke2012}.

\section*{Acknowledgements}
We gratefully acknowledge useful discussions with A. Rost and D. F. McMorrow. 
We would also like to thank G. Stenning and D. Nye for help with the instruments in the Materials Characterisation Laboratory at the ISIS Neutron and Muon Source, Kuang-Yu Samuel Chang and Roos Leenen for technical assistance with the DFT calculations, and Sebastian Bette for XRD characterizations. 
We acknowledge the support of the HFML-Radboud University (RU)/Netherlands Organisation for Scientific Research (NWO), a member of the European Magnetic Field Laboratory. This work is part of the research program ÒStrange MetalsÓ (Grant 16METL01) of the former Foundation for Fundamental Research on Matter, which is financially supported by the NWO and the European Research Council (ERC) under the European UnionÕs Horizon 2020 research and innovation programme (Grant Agreement No. 835279-Catch-22). 
We gratefully acknowledge support from the UK Engineering and Physical Sciences research council, grant EP/N034694/1. 
We acknowledge collaborative support from A.S. Gibbs, D. Fortes and the ISIS Crystallography Group for making available the 193Ir for the isotope work. Experiments at the ISIS Neutron and Muon Source were supported by a beamtime allocation RB1990395,  DOI:10.5286/ISIS.E.RB1990395,? from the Science and Technology Facilities Council.
The work of D. P. and V. M. was supported by Act 211 Government of the Russian Federation, contract 02.A03.21.0006.

\newcommand{\beginsupplement}{%
        \setcounter{table}{0}
        \renewcommand{\thetable}{S\arabic{table}}%
        \setcounter{figure}{0}
        \renewcommand{\thefigure}{S\arabic{figure}}%
     }
\beginsupplement

\end{document}